\documentclass[%
 reprint,
 amsmath,amssymb,
 aps,
 prl,
]{revtex4-1}

\usepackage{ulem}
\usepackage{color}
\usepackage{graphicx}% Include figure files
\usepackage{dcolumn}% Align table columns on decimal point
\usepackage{bm}% bold math
\usepackage{verbatim}

\begin{document}

%\preprint{}

\title{Self-Organized Velocity Pulses of Dense Colloidal Suspensions in Microchannel Flow}



\author{Philipp Kanehl}
\email[]{philipp.kanehl@campus.tu-berlin.de}
%\thanks{}
%\altaffiliation{Institute of Theoretical Physics, Technische Universit\"at Berlin, Hardenbergstr. 36, 10623 Berlin, Germany}

\author{Holger Stark}
\email[]{holger.stark@tu-berlin.de}
\affiliation{Institute of Theoretical Physics, Technische Universit\"at Berlin, Hardenbergstr. 36, 10623 Berlin, Germany}

\date{\today}

\begin{abstract}
We present a numerical study of dense colloidal suspensions in pressure-driven microchannel flow in two dimensions. The colloids are modeled as elastic and frictional spheres suspended in a Newtonian fluid, which we simulate using the method of multi-particle collision dynamics. The model reproduces periodic velocity and density pulse trains, traveling upstream in the microchannel, which are
found in experiments conducted by L. Isa \textit{et al.} [Phys. Rev. Lett. \textbf{102}, 058302 (2009)]. We
show that colloid-wall friction and the resultant force chains are crucial for the formation of these pulses. With increasing colloid density first solitary jams occur, which become periodic pulse trains at intermediate densities and unstable solitary pulses at high densities. We formulate a phenomenological continuum model and show how these spatio-temporal flow and density profiles
can be understood as homoclinic and periodic orbits in traveling-wave equations.
\end{abstract}

%\keywords{}

\maketitle


%introduction paragraph--------------------------------------------------------------------------

Understanding the collective dynamics of colloids in viscous fluids using microfluidic tools is an ongoing challenge \cite{Haw04, Roberts07, Bricard13, Lopez15, Ober15, Royer16, Peters16}. 
The complexity of collective phenomena contrasts with the simplicity of low-Reynolds-number flow, which is governed by the linear Stokes equation \cite{Happel12}. A prominent approach for their understanding is particle jamming, where dense colloidal flow in confinement is arrested due to the self-organized formation of  force chains \cite{Cates98}. The colloidal system thus becomes solid, although fragile with respect to small perturbations. Since jamming occurs in such diverse systems as granular matter\ \cite{Corwin05}, pedestrian and traffic flow \cite{Lighthill55, Henderson71, Helbing95}, but also, very prominently, during shear thickening of suspended corn starch \cite{Fall08, Brown14}, \citet{Liu98} suggested it as a universal principle governing dense particle systems.

It is well known that hydrodynamic lubrication prevents direct contact during collisions of micron-sized colloids suspended in a viscous fluid\ \cite{Arp77}. However, over the years research has suggested the necessity of direct particle contacts, for example,  due to surface roughness, to explain phenomena such as shear migration \cite{Phillips92, Cunha96, Kanehl15}, discontinuous shear thickening \cite{Brown12, Fernandez13}, and jamming \cite{Haw04}. Numerical studies by Seto {\it et al.} \cite{Seto13} and also work by Heussinger \cite{Heussinger13} and Fernandez \textit{et al.} \cite{Fernandez13} stressed the importance of implementing contact friction and thereby successfully modeled discontinuous shear thickening.

Pressure driven flow of dense colloidal suspensions through microchannels shows regular \cite{Isa09} and irregular \cite{Camp10} oscillations in flow speed, which have been attributed to the formation of transient jams.  Isa {\it et al.} \cite{Isa09} could indirectly verify the existence of density-rarefaction pulses traveling up-stream and stressed the importance of shear thickening under
confinement \cite{Fall08}.

In this article we combine two-dimensional simulations of a low-Reynolds-number fluid with a frictional contact model for particles \cite{Luding08}, in order to thoroughly study dense colloidal flow through microchannels. With increasing colloid density, we identify solitary jams, regular pulse 
trains, and solitary pulses in the colloidal flow similar to experimental observations\ \cite{Isa09, Camp10}. We stress the importance of colloid-wall friction and the formation of force chains for inducing a transition between free and jammed flow. This is the origin for traveling rarefaction pulses to occur. A newly formulated non-linear continuum model reproduces the flow instabilities from our simulation study.

%method paragraph--------------------------------------------------------------------------

\textit{Method:}
We use the mesoscale simulation method of multi-particle collision dynamics (MPCD) \cite{Male99, Male00, Pad06} to simulate the pressure driven flow of the dense colloidal suspension inside a channel of width $2w$ in two dimensions. MPCD has already successfully been applied to tackle diverse problems of soft matter physics by modeling the flow field around passive particles \cite{Prohm12, Pelto13, Singh14,Kanehl15} or active swimmers \cite{Zoettl14, Zoettl15, Aliza15, Heme15}. To solve the Navier-Stokes equations including thermal noise, the method generates a flow field using pointlike fluid particles, which perform alternating streaming and collision steps. 
The latter conserve translational and angular momentum as well as temperature using the Andersen thermostat for the collision rule \cite{Gomp09, Alla02}. For more details of the implementation we refer to our recent work \cite{Kanehl15}. The MPCD parameters are summarized in \cite{Supp_M}.

The colloidal suspension is modelled as a binary mixture of disks with neutral buoyancy and with radii $a/w=0.084$ and $0.118$. They have equal number and the size ratio $1.4$ in order to prevent crystallization \cite{Desmond09}. The frictional contacts of a colloid with the wall and other colloids are treated with a model commonly used in granular physics \cite{Cundall79, Luding08}.
When a colloid overlaps with another body, it experiences a repulsive spring force $\mathbf{F}_\mathrm{n}=-k_\mathrm{n}\boldsymbol{\delta}_\mathrm{n}$, where $k_\mathrm{n}$ is the spring constant. The vector $\boldsymbol{\delta}_\mathrm{n}$ measures the overlap distance and always points towards the collision partner along the normal vector $\mathbf{n}$ of both colliding surfaces. While in contact, the relative displacement $\boldsymbol{\delta}_\mathrm{t}$ in tangential direction $\mathbf{t} \perp \mathbf{n}$ (see \cite{Supp_M}) results in an elastic force $\mathbf{F}_\mathrm{t}=-k_\mathrm{t}\boldsymbol{\delta}_\mathrm{t}$ and torque $\mathbf{T}=a\mathbf{n}\times\mathbf{F}_\mathrm{t}$, where $k_\mathrm{t}$ is the tangential spring constant. However, when the static Coulomb friction force $|\mathbf{F}_\mathrm{t}|$ exceeds the maximum value $\mu|\mathbf{F}_\mathrm{n}|$, where $\mu$ is the friction coefficient, sticking is replaced by frictional sliding and the Coulomb friction force becomes $\mathbf{F}_\mathrm{t}=\mu|\mathbf{F}_\mathrm{n}|\mathbf{t}$ (see \cite{Luding08} for details). To implement stiff particles, we 
choose $k_\mathrm{n}/a \sigma = 800$ with $a=0.1w$ and $\sigma= -\nabla p$ is the pressure force acting on the fluid. 

In order to observe several velocity pulses in the channel, we choose a channel length $l/w=480$ and implement periodic boundary conditions in flow direction. Since the interesting variations of colloidal velocity $v$ and local area fraction $\phi$ occur along this direction, we average them over the channel cross section, $\psi(t, x)=\frac{1}{2w}\int_{-w}^w \psi(t, x,y)\, \mathrm{d}y$, with $\psi=\{v, \phi\}$. We determine $\phi$ using a Voronoi tessellation for the colloidal packing.

The high mean packing fractions $\overline{\phi} > 0.7$ are obtained by randomly placing up to 23000 initially pointlike colloids in the channel. Iteratively, they are expanded in size and overlaps are removed by molecular dynamics steps until the final colloidal size is reached \cite{Luba91}. The MPCD fluid is filled into the free space and each simulation runs on 6 processors using the parallel environment OpenMP. 

Densely packed particles experience high hydrodynamic friction quantified by the bulk viscosity. Therefore, all of our simulations are performed at channel Reynolds number $\mathrm{Re}\lesssim 0.1$ and in the high-P\'eclet number regime,
where any inertia and particle diffusion is negligible.

%figure 1 paragraph--------------------------------------------------------------------------

 \begin{figure}
 \includegraphics[width=0.97\linewidth]{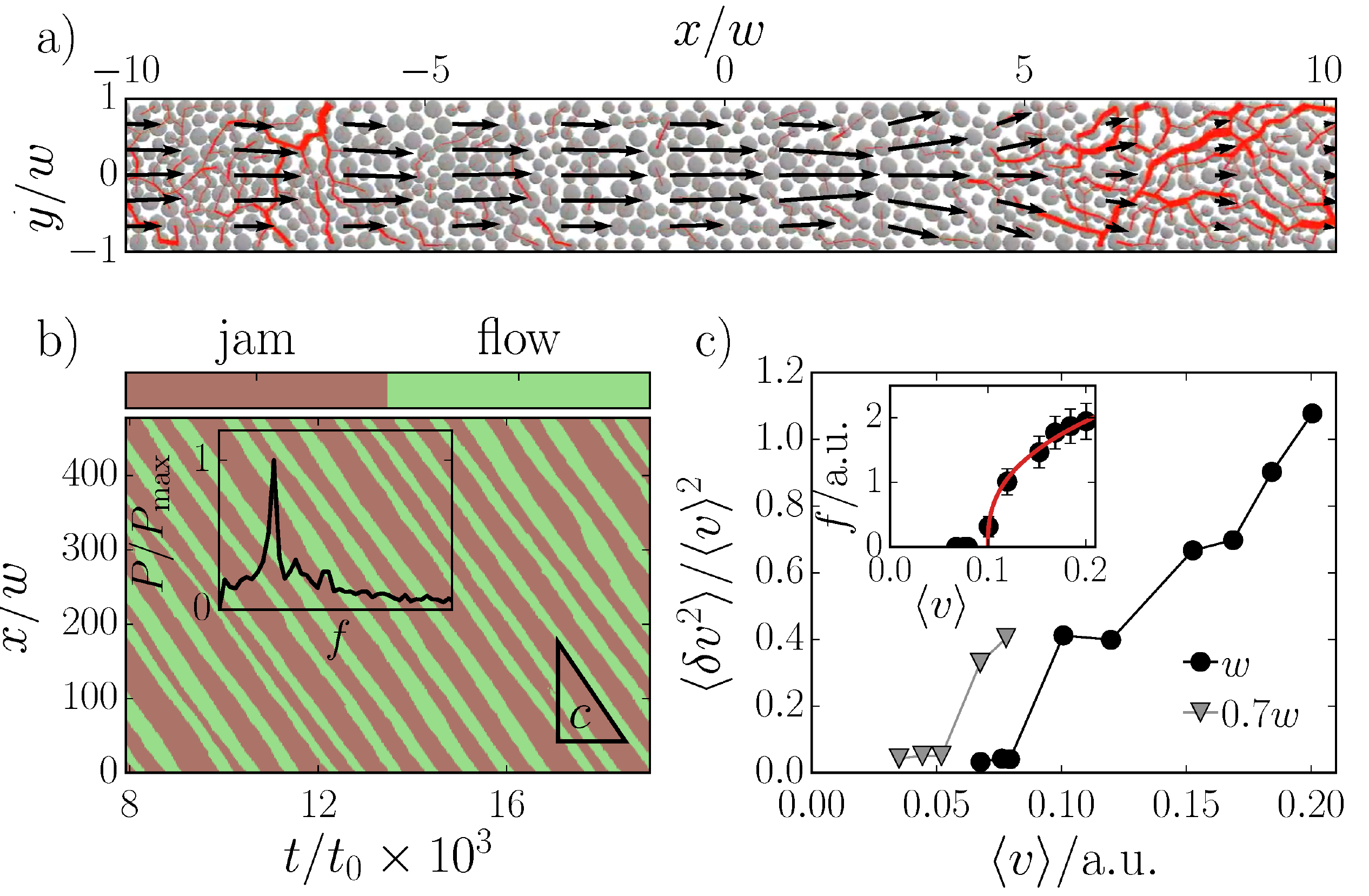}%
 \caption{Simulation at $\mu=1$ and $\overline{\phi}=0.75$. a) Snapshot of a velocity pulse inside the channel. Black arrows show the colloidal velocity field with the lateral component amplified by a factor 5. Red lines show the network of force chains.
b) Spatio-temporal representation of $v(x,t)$. The velocity pulses are indicated by flowing colloidal regions (green) separated by 
jammed regions (brown). $t_0=w/v_0$, where $v_0:= v(\phi{=}0)$ is the mean flow velocity without colloidal particles.
Inset: Power spectrum of $v(x,t)$ for fixed $x$. c) Normalized variance of $v(x,t)$ plotted versus mean flow velocity $\langle v\rangle$ for channels
of width $w$ and $0.7w$. Inset: Pulse frequency $f$ versus $\langle v\rangle$. Errorbars result from the finite channel length. The full red line
is a power-law fit with exponent 0.379. 
\label{fig:1}
}
 \end{figure}
 
\textit{Results:}
As soon as the pressure gradient is switched on, the suspension jams at different locations in the channel. Downstream of these jams, more dilute regions with increased
flow speed occur, whereas the jam grows upstream. Ultimately, regularly spaced rarefaction pulses develop and move upstream with speed $c$ (see video M1 \cite{Supp_M} in the lab and pulse frame, respectively). Fig.\ \ref{fig:1}a) shows such a rarefaction pulse with maximum flow speed at $x=0$ and roughly parabolic flow profile. Downstream, the flow diverges and at $x/w\approx 5$ a sharp transition to the jammed region occurs indicated by the red force chains spanning across the channel. Due to the higher packing density, the flow speed is low and the profile more flat. Upstream of the pulse ($x/w<-5$), colloids become gradually unjammed and accelerate with decreasing density.

Video M2 \cite{Supp_M} shows how several pulses self-organize into a stable regular pulse train from the start of the simulations. In Fig.\ \ref{fig:1} b), the resulting traveling pulses are then visualized as regularly spaced straight lines in the $x,t$ plane indicating free-flow and jammed regions. The pronounced peak of the power spectrum of $v(x,t)$ in the inset shows the regularity of the pulse train.

To quantify the strength of the traveling pulses, we introduce the variance of the flow field $\langle\delta v^2\rangle/\langle v\rangle^2$, 
where $\delta v=v-\langle v\rangle$ and $\langle ... \rangle$ means average over $x$ and $t$. Figure\ \ref{fig:1}c) shows that the pulse train emerges at a threshold value of $\langle v\rangle$, which we vary by the pressure force. The corresponding colloidal P\'eclet number is large:
$\mathrm{Pe} = \langle v \rangle a / D_0 \approx 100$, where $D_0$ is the diffusivity of a single colloid. The threshold value decreases for a more narrow channel.

We observe that the frequency of the pulses increase with $\langle v\rangle$, following
the power law $f \propto \left(\langle v\rangle-\langle v\rangle_\mathrm{thr}\right)^\beta$ with $\beta=0.38$ [inset of Fig.\ \ref{fig:1}c)].
This all agrees with the experimental findings of Isa {\it et al.} \cite{Isa09}.

%figure 2 paragraph--------------------------------------------------------------------------

 \begin{figure}
 \includegraphics[width=1.0\linewidth]{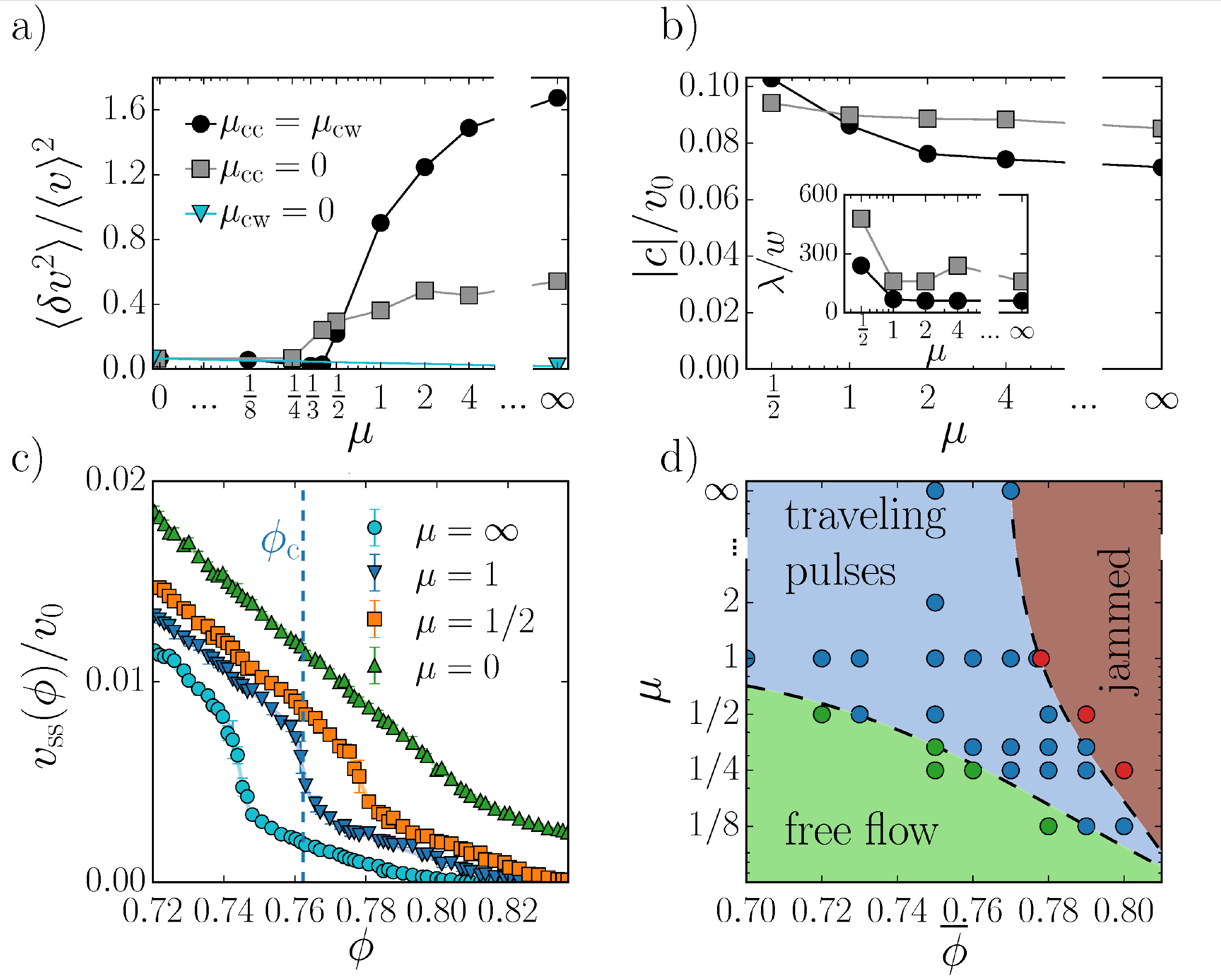}%
 \caption{a) Pulse strength $\langle\delta v^2\rangle/\langle v\rangle^2$ and b) pulse speed $c$ plotted versus friction coefficient $\mu$ for different friction compositions at $\overline{\phi}=0.75$. Inset: Distance $\lambda$ between pulses. Lines connecting the points are a guide to the eye. c) Steady-state flow velocity $v_\mathrm{ss}$ versus density $\phi$ for different $\mu$ in short channels. Errorbars indicate standard deviations from five simulation runs. d) State diagram $\mu$ versus mean density $\overline{\phi}$ showing jammed, traveling pulses, and free-flow regions. Dots indicate conducted simulations. \label{fig:2}}
\end{figure}

Similar to discontinuous shear thickening in hard-sphere suspensions, the implemented Coulomb friction plays a decisive role in observing the regular pulse trains. In Fig. \ref{fig:2}a) we plot the pulse strength versus friction coefficient. In the first case, friction between colloids ($\mu_\mathrm{cc}$) and friction between colloids and wall ($\mu_\mathrm{cw}$) are equal to $\mu$. In a continuous
transition, velocity pulses emerge at $\mu \approx 0.5$ and the pulse strength sharply increases. 
Note that this increase in $\langle\delta v^2\rangle/\langle v\rangle^2$ is mainly due to the fact that the distance $\lambda$ between the traveling pulses sharply decreases as indicated in the inset of Fig.\ \ref{fig:2}b), while the pulse height remains approximately the same. If the friction between the colloids is switched off, $\mu_\mathrm{cc}=0$, the distance $\lambda$ is larger [inset of Fig.\ \ref{fig:2}b)], which explains the overall smaller pulse strength in Fig.\ \ref{fig:2}a).
However, more importantly for frictionless walls, $\mu_\mathrm{cw}=0$, traveling pulses do not occur even if colloidal friction is present. Force chains spanning across the channel cannot form and thus the jammed regions, essential for the pulses to occur, cannot develop. In Fig.\ \ref{fig:2}b) the pulse speed $c$ decreases close to the transition and then stays approximately constant. A similar behavior is observed for the distance $\lambda$ between the pulses [see inset of Fig.\ \ref{fig:2}b)].

 \begin{figure}
 \includegraphics[width=0.97\linewidth]{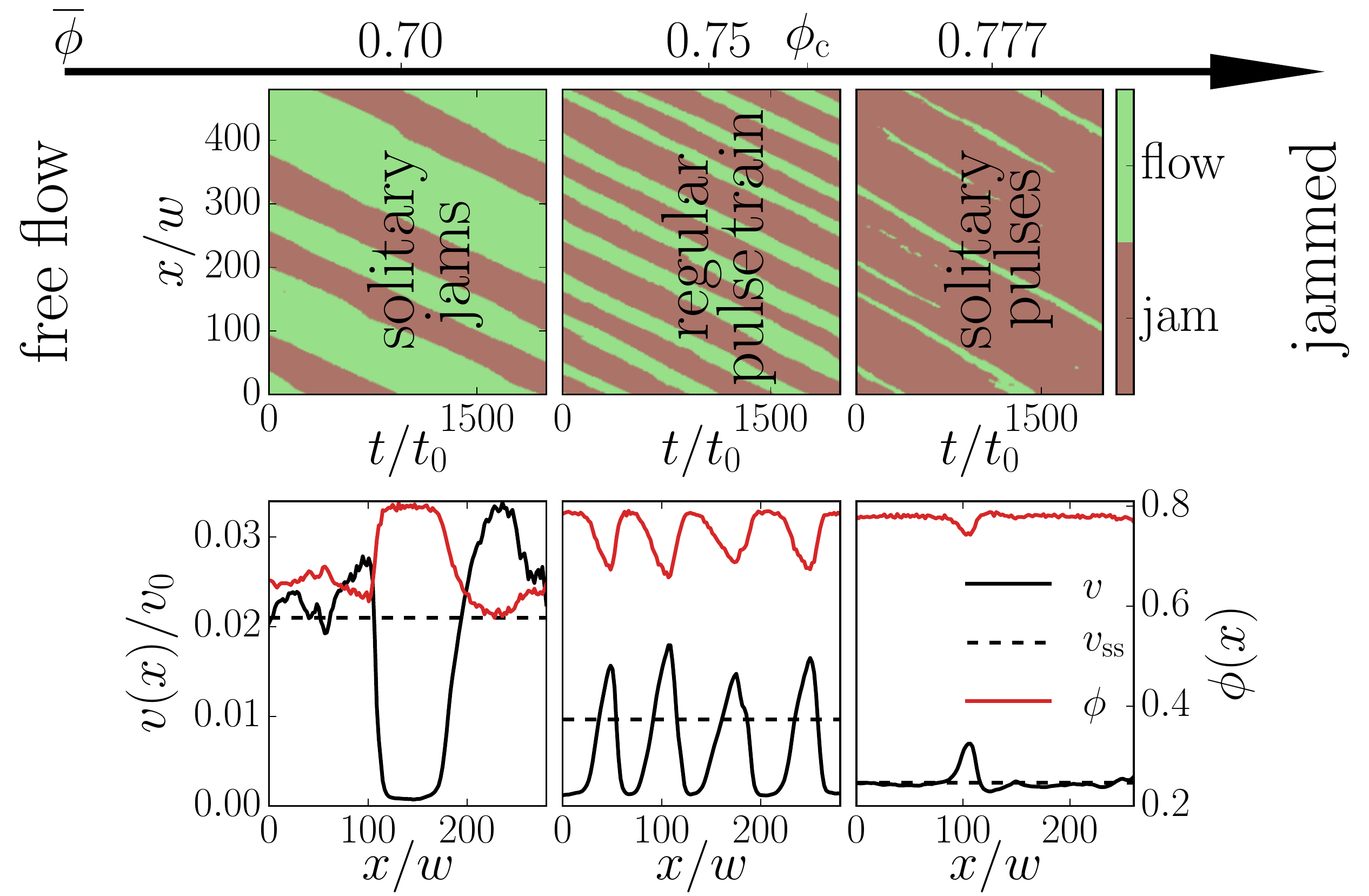}%
 \caption{Spatio-temporal representation (top) and snapshots (bottom) of flow velocity $v(x,t)$ and area fraction $\phi$ for different mean densities $\overline{\phi}$ at $\mu=1$. The dashed lines indicate the steady-state velocity $v_\mathrm{ss}$ at the respective $\overline{\phi}$. \label{fig:3}}
 \end{figure}

In simulations with sufficiently short channels, the colloidal flow remains uniform. Figure\ \ref{fig:2}c) plots the flow velocity $v_\mathrm{ss}$ in this steady state versus packing fraction $\phi$ for different friction coefficients $\mu$. While for zero friction, $\mu=0$, $v_\mathrm{ss}$ decreases smoothly towards zero, for finite friction, $\mu>0$, a jump of $v_\mathrm{ss}$ occurs at a critical density $\phi_\mathrm{c}$ and one can clearly distinguish between states of free flow and jammed flow. This feature can qualitatively explain the formation of traveling pulses \cite{press_grad}. 
The colloidal density in the free-flow regime fluctuates locally about its mean value. If such fluctuations exceed $\phi_\mathrm{c}$, the colloidal flow strongly slows down. Frictional contacts generate force chains, which further arrest the flow as more colloids accumulate upstream of the jam. Downstream, the suspension rarefies as colloids still flow freely. This rarefaction dissolves the force chains and the jammed region moves upstream.

The state diagram in Fig.\ \ref{fig:2}d) summarizes the behavior in long channels. The colloidal flow is classified as traveling pulses if $\langle\delta v^2\rangle/\langle v\rangle^2>0.1$; otherwise, we refer to it as free flow for mean velocity $\langle v\rangle/v_0\geq 0.005$ or jammed flow if $\langle v\rangle/v_0< 0.005$. The region of traveling pulses becomes narrower and shifts to higher densities as the friction coefficient decreases. Larger densities are needed to build up force chains via frictional contacts, while rarefaction pulses cannot develop if the density is too high.

%figure 3 paragraph--------------------------------------------------------------------------

Different flow patterns of traveling pulses occur for different mean densities $\overline{\phi}$ as illustrated in Fig.\ \ref{fig:3}. At lower densities solitary jams develop. They persist throughout the simulation and the distance between the jammed regions may vary largely. Stable and regular trains of pulses form close to the critical density $\overline{\phi}\approx\phi_\mathrm{c}$. Finally, near the boundary to the jammed flow in Fig.\ \ref{fig:2}d) only weak solitary pulses of finite life time appear throughout the channel as evident by the green stripes, which start or end in the jammed flow region. Interestingly, \citet{Camp10} made a similar observation in an experiment on dense colloidal flow into a channel with converging cross section. They found that by slightly increasing the colloidal volume fraction, periodic oscillations in flow speed measured at the channel inlet transformed into transient pulses separated by irregular long-lived jams.

%theory and figure 4 paragraph-------------------------------------------------------------
\textit{Continuum model:}
In the following we present a phenomenological theory, which is able to explain our numerical results. It is motivated by continuum traffic-flow models that capture the formation of shock fronts \cite{Lighthill55}, density autosolitions, and periodic density modulations \cite{Payne79, Kerner93, Kerner94}. The colloidal area fraction $\phi$ obeys the one-dimensional continuity equation:
\begin{equation} 
\frac{\partial\phi}{\partial t}+\frac{\partial v\phi}{\partial x} = 0 \, ,\label{eq:conserve}
\end{equation}
where we neglect diffusion due to the high P\'eclet number.

For the colloidal flow velocity we formulate a one-dimensional phenomenological equation. First,
we refer to the steady-state velocity $v_{\mathrm{ss}}$ plotted in Fig.\ \ref{fig:2}c) 
and extract a density-dependent friction coefficient $\xi(\phi)=\sigma / v_\mathrm{ss}(\phi)$ that originates from shear stresses due to the non-uniform colloidal flow profile in the channel cross section
and colloidal contact friction. Note that $\xi(\phi)$ was determined in short channels, where colloidal flow was not varying along the channel axis. However, for traveling pulses the colloidal flow velocity is
non-uniform along the channel axis, which gives rise to momentum diffusion with a current $j \propto \partial v / \partial x$. Adding both contributions and neglecting inertia, we arrive at
\begin{equation} 
0 =  \sigma-\xi(\phi) v - \frac{\partial j}{\partial x} =
\sigma-\xi(\phi) v+\nu \frac{\partial^2 v}{\partial x^2},
\label{eq:newton}
\end{equation}
where $\nu$ is an effective viscosity \cite{visc}. For $\xi(\phi) = \sigma / v_\mathrm{ss}(\phi)$ we use an analytical fitting function, which we obtain by fitting the data for $v_\mathrm{ss}(\phi)$ in Fig.\ \ref{fig:2}c). The fitting function $v^\mathrm{fit}_\mathrm{ss}(\phi)$ approximates the characteristic jump at 
$\phi_\mathrm{c}$ by $\tanh[(\phi_\mathrm{c}-\phi) / \Delta \phi]$, where $\Delta \phi$ controls its width. Furthermore, $v^\mathrm{fit}_\mathrm{ss}(\phi)$ describes the overall decline of 
$v_\mathrm{ss}(\phi)$ by a linear and quadratic function on the left and right of $\phi_\mathrm{c}$, respectively (for further details 
we refer to \cite{Supp_M}). Finally, rescaling lengths by $w$, velocity by $V=\sigma/\xi_\mathrm{c}$ with $\xi_\mathrm{c}=\xi(\phi_\mathrm{c})$, and introducing reduced quantities ($x/w\to x$, $v/V\to v$, $tV/w\to t$, and $\xi/\xi_\mathrm{c}\to \xi$), we write Eqs.\ (\ref{eq:conserve}) and (\ref{eq:newton}) in dimensionless form with $\sigma \rightarrow 1$ and $\nu / (\xi_\mathrm{c}w^2) \rightarrow K$.

\begin{figure}
\includegraphics[width=1.0\linewidth]{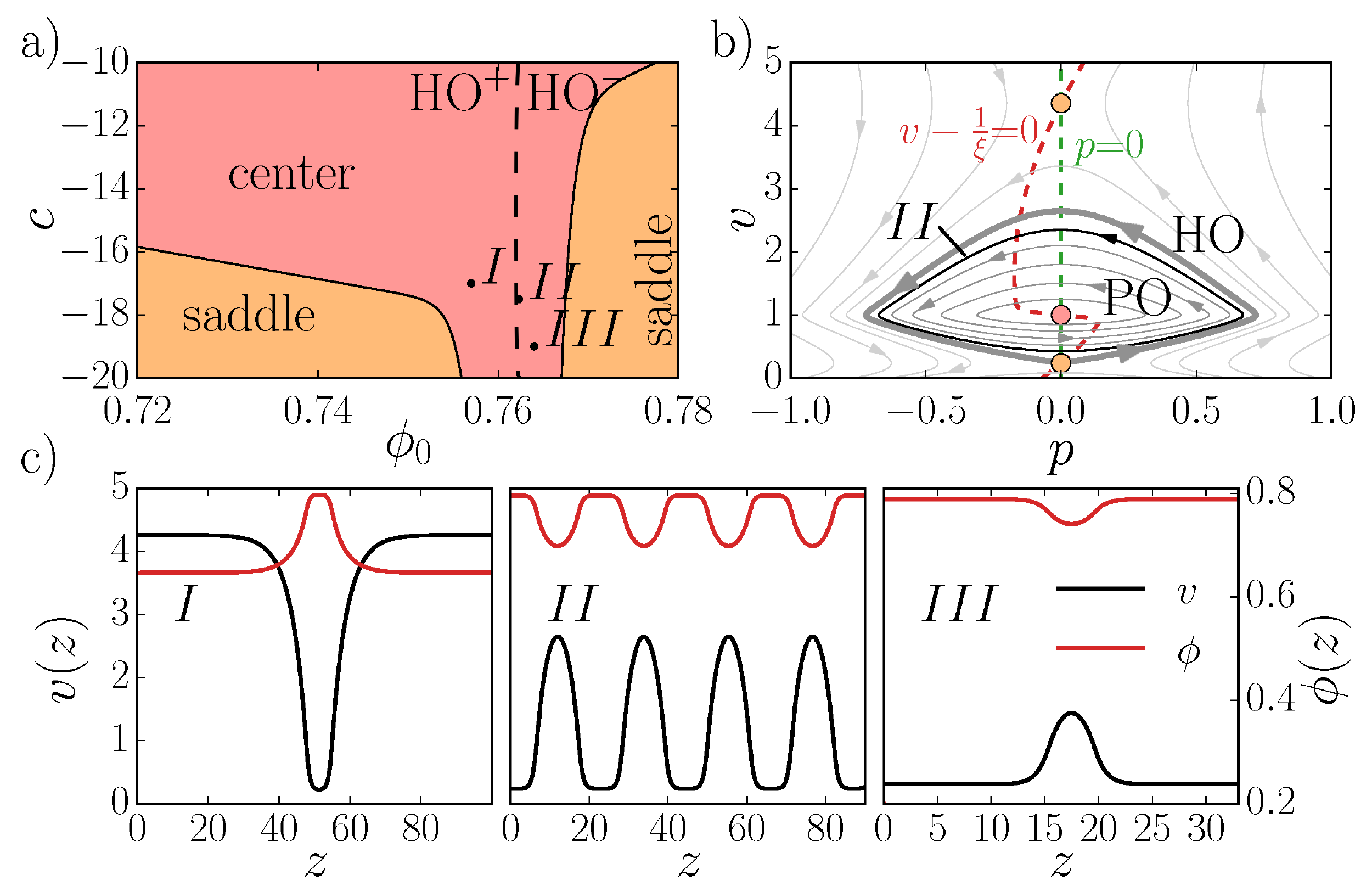}%
\caption{a) Classification of fixed points in the $c,\phi_0$ plane for Eqs.\ (\ref{eq:twe}) for $\mu=1$. The homoclinic orbits HO$^{+/-}$ touch the free-flow/ jammed-flow saddles. b) Phase portrait for $q=13.72$ and $c=-17$ showing period orbits (POs) enclosed by a HO, which touches the jammed-flow saddle. The PO indicated by $\uppercase\expandafter{\romannumeral 2}$ generates the periodic pulse train in c).
The three fixed points are classified by the color code of a). c) Flow ($v$) and density ($\phi$) profiles generated by closed orbits in the phase portrait for parameters marked with $\uppercase\expandafter{\romannumeral 1}$-$\uppercase\expandafter{\romannumeral 3}$ in a). 
\label{fig:4} }
\end{figure}

We search for traveling-wave solutions of the form $\phi(z)$ and $v(z)$ with $z=(x-ct)/\sqrt{K}$. Using this ansatz in the 
dynamic equations for $v$, $\phi$ and integrating Eq.\ (\ref{eq:conserve}) once, we obtain
\begin{equation} 
\phi(v)=\frac{q}{v-c} \quad \mathrm{and} \quad
0 = 1 - \xi(\phi(v)) v + \frac{\partial^2 v}{\partial z^2} \, .
\label{eq:twe0}
\end{equation}
Here the constant of integration $q$ is the current density in the co-moving frame.

The first relation in Eqs.\ (\ref{eq:twe0}) holds true, in particular, within a jam, where $\phi_\mathrm{j}>\phi_\mathrm{c}$ and the particles creep with a velocity $v_\mathrm{j}$, as well as at the critical density $\phi_c$, where particles move with $v_\mathrm{ss}(\phi_\mathrm{c})$. Thus, by eliminating $q$, we find an expression for the pulse speed $c$:
\begin{equation} 
c=\frac{\phi_\mathrm{j}v_\mathrm{j}-\phi_\mathrm{c}v_\mathrm{ss}(\phi_\mathrm{c})}{\phi_\mathrm{j}-\phi_\mathrm{c}}\approx-\frac{\phi_\mathrm{c}}{\phi_\mathrm{j}-\phi_\mathrm{c}}v_\mathrm{ss}(\phi_\mathrm{c}) \, . 
	\label{eq:wavespeed}
\end{equation}
To arrive at the second equation, we used $v_\mathrm{j} \ll v_{\mathrm{ss}}$ and $\phi_\mathrm{j}  \approx \phi_\mathrm{c}$. 
Now, since $0 < (\phi_\mathrm{j}-\phi_\mathrm{c}) / \phi_\mathrm{c} \ll 1$ in our simulations, Eq.\ (\ref{eq:wavespeed}) confirms our findings that the pulse speed $c$ is negative and that the pulse always travels against the colloidal flow with a speed much larger than $v_\mathrm{ss}(\phi_\mathrm{c})$.

To identify the different type of rarefaction pulses illustrated in Fig. \ref{fig:3}, we convert the second relation of
Eqs.\ (\ref{eq:twe0}) into two first-order ordinary differential equations:
\begin{equation}
\frac{\mathrm{d} v}{\mathrm{d} z}=p
\enspace \mathrm{and} \enspace \frac{\mathrm{d} p}{\mathrm{d} z} = \xi(\phi(v))v -1 \, .
\label{eq:twe}
\end{equation}
The theory of dynamical systems helps to analyze Eqs.\ (\ref{eq:twe}). Their fixed points $(p_0,v_0)$ follow
from the intersections of the two nullclines $\mathrm{d}p/\mathrm{d}z=0$ and $\mathrm{d}v/\mathrm{d}z=0$, 
\begin{equation}
p_0=0 \quad \mathrm{and} \quad
 v_0 = 1 / \xi(\phi(v_0))
\, .
\label{eq:ncl}
\end{equation}
Together with the first relation of Eqs.\ (\ref{eq:twe0}), $\phi = q/ (v-c)$, they determine
a uniform density and channel flow. In the supplemental material\ \cite{Supp_M} we identify the possible types of the fixed points as saddle and centers.
The close-up diagram in Fig.\ \ref{fig:4}a), which we use in the following discussion, classifies the different types of fixed points in the $c,\phi_0$ plane. A larger view of it is presented in \cite{Supp_M}.

The phase portrait in Fig.\ \ref{fig:4}b) for specific values of $q,c$ shows flow lines, which correspond to solutions of the full non-linear equations (\ref{eq:twe}). Two saddle points (ochre dots) located on the respective free-flow and jammed-flow branch of  $v_\mathrm{ss}(\phi)$ surround a  center (red dot). 
In general, flow lines leave saddle fixed points from $z=-\infty$ and diverge to infinite velocities at $z =+\infty$ or
vice versa. Therefore, they do not give finite flow and density profiles along the microchannel. 
In contrast, a center fixed point is surrounded by periodic orbits (POs),
the amplitude and period of which increase with growing distance from the center. We indicate the
PO, which produces the periodic pulse train $\uppercase\expandafter{\romannumeral 2}$ in Fig.\ \ref{fig:4}c).
The corresponding density profile is determined from $\phi = q/ (v-c)$. When a PO touches
a saddle point, it becomes a homoclinic orbit (HO). The HO connected to the jammed-flow saddle [see Fig.\ \ref{fig:4}b) at small $v$] generates a solitary pulse [type $\uppercase\expandafter{\romannumeral 3}$ in Fig.\ \ref{fig:4}c)], while a collision with the free-flow saddle (large $v$) gives a solitary jam (type $\uppercase\expandafter{\romannumeral 1}$).
Thus all three finite solutions of our phenomenological model nicely reproduce the characteristic traveling flow and density profiles in Fig.\ \ref{fig:3}, which we obtained in the simulations.

To conclude, our numerical model reproduces all features of the experimental counterpart including traveling pulse trains or rarefaction pulses moving upstream. It identifies a transition between free and jammed flow due to force chains spanning the channel cross section as the crucial reason for the formation of flow instabilities. They include solitary jams, regular pulse trains, and solitary pulses, which set in with increasing density, wall-colloid friction, and flow velocity. We are also able to observe these rarefaction pulses in three-dimensional simulations as video M3 in \cite{Supp_M} demonstrates. A nonlinear phenomenological model, which we analyzed with methods from nonlinear dynamics to identify traveling-wave solutions, is able to describe all the traveling pulse profiles from our simulations.

Our study is a good example, how the unifying idea of jamming determines the complex flow of dense colloidal systems. It suggest that repulsive walls and that reducing colloid-wall friction, packing fraction, or flow speed prevent colloidal jamming. These insights may help in addressing unwanted channel or nozzle clogging in industrial processing \cite{Lewis02, Zuriguel14}

\begin{acknowledgments}
We thank H. Engel, L. Isa, C. Lozano, and T. P\"oschel for helpful discussions. 
We also appreciate very helpful comments from the referees.
This work was supported by the Deutsche Forschungsgemeinschaft (DFG) through the research training 
group GRK 1558 and within the project STA/10-1.

\end{acknowledgments}


\normalem
\bibliography{bibfile}

\begin{thebibliography}{50}
\bibitem [{vis()}]{visc}%
  \BibitemOpen
  \href@noop {} {}\bibinfo {howpublished} {For simplicity, we neglect here any
  dependence of $\nu$ on density $\phi$.}\BibitemShut {Stop}%
\end{thebibliography}

\end{document}